\documentclass[aps,prl,twocolumn,floats]{revtex4}
\usepackage{graphicx}

\begin{document}

\title{Fano effect  of strongly interacting
 quantum dot in contact with superconductor}
\author{ Anatoly Golub $^1$ and Yshai Avishai $^{1,2}$ }

\address{ $^1$ Department of Physics,
Ben-Gurion University of the Negev,
          Beer-Sheva,
         Israel \\
         $^2$ and Ilse Katz Center for 
Nano-Technology, Ben-Gurion University of the
         Negev,
          Beer-Sheva,
         Israel}


\begin{abstract}
The physics of a system consisting of an Aharonov Bohm (AB)
interferometer with a single level interacting quantum dot (QD) on
one of its arms, and attached to normal (N) and superconducting
(S) leads is elucidated. Here the focus is directed mainly on
N-AB-S junctions but the theory is capable of studying S-AB-S
junctions as well. The interesting physics emerges under the
conditions that both the Kondo effect in the QD and the the Fano
effect are equally important. As the Fano effect becomes more
dominant, the conductance of the junction is suppressed.
\end{abstract}
\maketitle

 {\it Motivation}: Transport through a mesoscopic Aharonov-Bohm
 (AB) ring (with an interacting quantum dot (QD) situated on one of its arms)
  weakly attached to normal (N) metallic leads is the subject of recent
  intensive experimental \cite{yacoby,weil,Kobayashi} and theoretical
   \cite{oreg,gefen,hof,bulka,david,aha}
  studies. To the above class of experiments
  one may also add STM measurements
  on a single magnetic atom adsorbed on a gold surface \cite{Madhavan}.
  The main result of the theoretical analysis \cite{hof,bulka}
  indicates that in such N-AB-N junctions there is an interplay between
  two fundamental physical phenomena, namely,
  the Fano and the Kondo effects. The
  Fano effect is related to an interference
  between two electron waves one passing through the quantum dot (with a
  discrete level) and one travelling along the
  direct 'reference' channel characterized by
  its continuous spectrum. It results in an asymmetric shape of
  the conductance as function of the applied bias (or gate) voltage.
  The Kondo effect is one of the simplest
manifestations of many-body physics exhibiting strong
correlations. It plays an important role in electron transport
through quantum dots, where the role of a magnetic impurity is
played here by the presence of localized electrons
 \cite{glazman88,ournca}.
 The interplay between these two effects causes the suppression of
 Kondo plateau with increasing transmission through the direct
 channel.

\begin{figure}
\begin{center}
\includegraphics [scale=0.45 ]{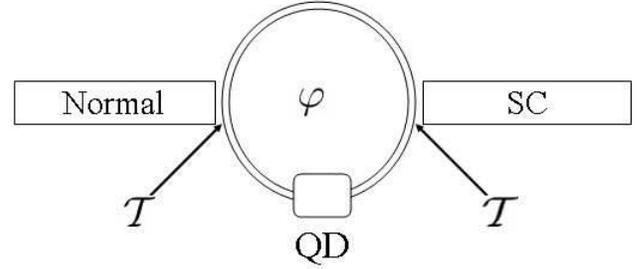}
\end{center}
\caption {\label{fig1} Schematic form of an N-AB-S junction.}
\end{figure}

An analysis of the conductance of an AB interferometer
  when one of the leads
 is a normal metal and the other one is a
 superconductor (S) is the subject of the present research. The
 geometry of such N-AB-S junction is schematically displayed in
 Fig.\ref{fig1}.
 We derive a general formula for the conductance of N-AB-S and S-AB-S
 junctions.
 \\
\hspace{10mm}
As for the motivation,  there is indeed a growing
interest
 in these types of superconducting junctions,
especially in the light of recent
experiments
 which involve carbon nanotubes as weak links\cite{bu,Kas,lin}. They are
 characterized by relatively high Kondo temperature which helps the
 formation of Kondo plateau in conductance experiments.
 The Kondo finger-prints in such
 junctions is expected to survive even for cases where
  the superconducting gap approaches the Kondo temperature from below,
  that is, $\Delta\sim T_{K}$ (see Refs.\cite{ambe,raimondi,us1}).

{\it Model Hamiltonian and the current}: In the system under
consideration (see Fig. \ref{fig1}), transport of an electron
between the two-dimensional electrodes (N on the left and S on the right) is
possible via two paths, either through the QD,
or else through the other 'direct' channel.
The dynamics of the system is governed by the Hamiltonian
\begin{eqnarray}
&& H=H_{L}+H_{R}+H_{d}+H_{t}+H_{LR}, \label{H}
\end{eqnarray}
in which $H_{j}$, ($j=L,R$) are the  Hamiltonians of the electrons
in the electrodes which depend on the electron field operators
$\psi_{j\sigma}({\bf r},t)$ where ${\bf r}=(x,y)$ and $\sigma=\pm$
is the spin index,
\begin{eqnarray}
\mbox{H}_j=\int dr [\psi^{\dagger}_{j \sigma} (\mbox{r})\xi({\bf
\nabla})\psi_{j \sigma}
        -\alpha \psi^{\dagger}_{j
\uparrow}(\mbox{r})\psi^{\dagger}_{j \downarrow}
        \psi_{j \downarrow}(\mbox{r})\psi_{j \uparrow}(\mbox{r})].
\label{BCS}
\end{eqnarray}
Here $\alpha$ is the BCS coupling constant and $\xi ({\bf
\nabla})=-{\bf \nabla}^2/2m-\mu$ with $\mu$ being the chemical
potential at temperature $T$. The dot is represented by a single
level Anderson impurity with energy $\epsilon_{0}<0$ and Hubbard
repulsion $U$. Later on we will concentrate on the Kondo
regime by setting $U \to \infty$ and assuming that
$|\epsilon_{0}|$ exceeds any other energy scale except $U$. The
Nambu representation for all fermion operators
($\psi_{\sigma}({\bf r})$ for the lead electrons and $c_{\sigma}$
for the QD electrons) is employed below (for example $
c^{\dagger}\equiv(c_{1}^{\dagger},c_{2}^{\dagger})
=(c_{\uparrow}^{\dagger},c_{\downarrow})$). The dot Hamiltonian
then reads,
\begin{eqnarray}
&& H_{d}=\epsilon c^{\dagger}\tau_{z} c +U n_{\uparrow}n_{
\downarrow}, \label{HD}
\end{eqnarray}
  where $n_{\sigma}=c_{\sigma}^{\dagger}c_{\sigma}$ and the
 Pauli matrix $\tau_{z}$ acts in Nambu space. Assuming a symmetric
 junction, the tunnelling Hamiltonian $H_{t}$ from the leads to the dot is,
\begin{eqnarray}
&&   H_{t}= {\cal T } \sum_{j=L,R}
 c^{\dagger} \tau_{z} \psi_{j }({\bf0})+h.c.,
  \label{Ht}
\end{eqnarray}
 where ${\cal T}$ is the tunnelling amplitude.
The occurrence of direct transport channel is represented
 by the term
 $H_{LR}$  between left and right leads. The gauge is chosen such
 that the AB flux appears only in this direct term,
\begin{eqnarray}
&& H_{LR}={\cal W}\psi_{R}^{\dagger}({\bf 0})
\tau_{z}F\psi_{L}({\bf 0})+h.c.,  \label{HLR}
\end{eqnarray}
where $F=\exp(i\varphi \tau_{z})$  is the AB phase factor. Note
that unlike the case of N-AB-N junctions, the phase factor is {\it
non-Abelian}.

 In the mean field slave boson approximation (MFA) which is employed below to
 handle the Kondo problem (in the limit $U \to \infty$), the tunnelling
 amplitude is modified by replacing the (slave) boson operators
 by their mean-field values,
 ($b \rightarrow <b>$)
 and the Hubbard repulsion term is dropped.
 This procedure imposes a
constraint on the number of bosons and fermions and the Hamiltonian
must now include a term which prevents double
occupancy in the limit $U\rightarrow\infty$,
\begin{eqnarray}
&& H_{c}=(\epsilon_ {r}-\epsilon)(c^{\dagger}\tau_{z}
c+b^{\dagger}b-1), \label{Hc}
\end{eqnarray}
where $\epsilon_ {r}$ is the renormalized new level position which
has the role of a Lagrange multiplier.

Formula for the current at one lead, say the right
(supreconducting), is derived from the basic relation
$I=-ie<[n_{R},H]> $. It is convenient to display the result of the
commutation relation in terms of Keldysh Green's functions
(GF) , namely,
\begin{eqnarray}
I&=&\frac{ e}{h}[\sqrt{\gamma}Tr(F^{\dagger}G_{LR}^{K}+H.c)+
\pi{\cal T} \rho Tr(G_{dR}^{K}+H.c)],
 \label{cur}
\end{eqnarray}
where $\rho$ is the density of states in the leads (assumed to be
the same in both), $\gamma=\pi^2 {\cal W}^2 \rho^2 $ and the trace
operation acts in Nambu and energy spaces ($Tr[A(\omega)]\equiv
\sum_{i} \int A_{ii}(\omega) d \omega$). Other notations are
$G_{LR}^{K}=G_{LR}^{<}+G_{LR}^{>}$ (similarly for $G_{dR}^{K}$),
where $G_{LR}^{<}(t)=i< \psi_{R}^{\dagger}\psi_{L}(t)>$ and
$G_{dR}^{<}(t)=i< \psi_{R}^{\dagger}c(t)>$.

Employing equation of motion method for all Keldysh
GF it is possible to express the current in terms of the
dot GF. We single out the current through the
direct channel by writing $G_{LR}^{K}=G_{LdR}^{K}+G_{dir}^{K}$,
with,
\begin{eqnarray}
G_{dir}^{K}&=&2\sqrt{\gamma}[g_{LL}F\tau_{z}g_{R}]^{K} \\
 G_{LdR}^{K}&=&\Gamma
[g_{LL}(1+2\sqrt{\gamma}F\tau_{z}g_{R}){\hat{G}}(1+2\sqrt{\gamma}g_{L}F\tau_{z})
g_{RR}]^{K}\nonumber\\
G_{dR}^{K}&=&{\cal{T}}
[\tau_{z}{\hat{G}}(1+2\sqrt{\gamma}g_{L}F\tau_{z})g_{RR}]^{K}.
\label{G}
\end{eqnarray}
Here $\Gamma=2\pi \rho {\cal{T}}^2 $ stands for tunneling rate
through the dot, ${\hat{G}}=\tau_{z}G\tau_{z}$, $G$ is the dot
 GF, $g_{L,R}$ are generally nonequilibrium GF of the leads
(left,right). The other two GFs include multiple scattering events
and have the form,
\begin{eqnarray}
g_{LL}&=&(1-4\gamma g_{L}F\tau_{z}g_{R}\tau_{z}F^{\dagger})^{-1}g_{L}\nonumber\\
g_{RR}&=&(1-4\gamma
g_{R}F^{\dagger}\tau_{z}g_{L}\tau_{z}F)^{-1}g_{R}.
 \label{gg}
\end{eqnarray}
Each GF ($ g_{L,R},g_{LL}, g_{RR}$, and $G$) has a standard $2
\times 2$ Keldysh matrix structure with elements $ G_{11}=G^{R}$,
$G_{22}=G^{A}$, $G_{21}=0$, $G_{12}=G^{K}$ . The superscript $K$
stands for Keldysh component of the matrix product occurring in
the square brackets of Eq.(7-\ref{G}). The set of equations
(7-\ref{gg}) together with the expression (\ref{cur}) for the
current formally complete our task . They give an expression for
the current through a closed AB interferometer in terms of the GF
of the (strongly interacting) QD. The above formalism has been
tested for the case of linear conductance with normal leads on
both sides (N-AB-N junction) and the result agrees with pertinent
calculations \cite{hof}.

These rather general expressions are now employed for elucidating
a particular case of interest, namely, the zero temperature limit
of the linear conductance in N-AB-S junctions. Specifically, as in
Fig. (\ref {fig1}) the left lead is a normal metal biased with an
external voltage V whose GFs are, $ g_{L}^{R,A}=-\pm i/2$,
   $ g_{L}^{K}=-i(\tanh(\omega/2T)+\delta f \tau_{z}) $;
   $ \delta f=eV/2T(\cosh(\omega/2T))^{-2} $.
On the other hand, the right lead is an unbiased (s-wave)
superconductor with gap $\Delta$ whose equilibrium GFs have a
standard form\cite{us1}.

 {\it Mean Field Approximation and Conductance}:
Within the MFA the algorithm starts with calculations pertaining
to a geometrically identical system albeit with {\it noninteracting} QD
$(U \to 0)$. Then, at the end, the replacement $\Gamma \rightarrow
T_{k},\epsilon \rightarrow \epsilon_ {r}$ is executed in the
conductance formula. These two quantities (the Kondo temperature
$T_{K}=\Gamma b^2 $ and the effective renormalized position of the
level) are evaluated by solving two mean field equations similar to
those derived in Refs. \cite{raimondi,us1}.
At zero bias one is free to adopt the Matsubara
form of these equations. The first one, $2 \pi (\epsilon_{
r}-\epsilon)+\Gamma Tr({\cal{G}}\Lambda)=0$, follows from the
extremum requirement of the effective action over $\epsilon_ {r}$. The second
one, $2 \pi T_{K}+ \Gamma Tr({\cal{G}}\tau_{z})=0$, reflects the
single occupation condition. Here
\begin{eqnarray}
&& \Lambda(\omega)=i\omega-\epsilon_ {r}\tau_{z}
+(\tau_{z}-i\sqrt{\gamma}\frac{\omega}{|\omega|}F){\cal{G}}
_{RR}(\tau_{z}- i\sqrt{\gamma}\frac{\omega}{|\omega|}F^{\dagger})
 \nonumber \\
&& {\cal{G}}^{-1}(\omega)= i\omega-\epsilon
_{r}\tau_{z}-T_{K}\Lambda,
 \label{Gm}\\
&& {\cal{G}}
_{RR}(\omega)=-\frac{i}{2d(\omega)}(\omega+\gamma\frac{\omega}{|\omega|}\sqrt
{\omega^ {2}+\Delta^{2}}-i\tau_{x}\Delta) \label{grrm},
\end{eqnarray}
 where $d(\omega)=(1 + \gamma^2))\omega^2 \sqrt{\Delta ^2
+\omega^2} + 2\gamma \omega^2|\omega| $.
 Expressions (\ref{Gm},\ref{grrm}) for the dot Matsubara GF were
 obtained from the effective action derived from the
 Hamiltonian of the N-AB-S system.
The self-consistence equations were solved  for different values
of the direct transmission parameter $\gamma$ while the level
positions $\epsilon$ are scanned over the wide range (-0.4,0.1).
The parameters are tuned such that the carbon nanotubes in Kondo
regime considered in reference \cite {bu} can be satisfactorily
regarded as an Anderson impurity. The half width of the normal
electrode density of states is served as an energy unit and the
tunnelling rate through the dot is fixed at $\Gamma=0.15$. The
superconducting gap $\Delta$ is set at a value $\Delta=0.03$. For
the majority of level positions $\epsilon$ the $\Delta \leq T_{K}$
and thus the MFA is hence justified.

\begin{figure}[ht]
\begin{center}
\includegraphics [width=0.45\textwidth ]{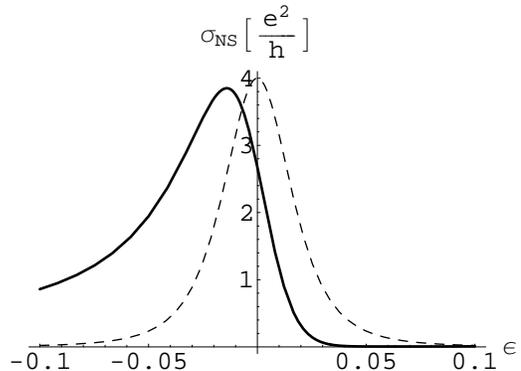}
\caption {\label{fig20} Linear conductance of an N-AB-S junction
(see Fig. \ref{fig1}) at $T=0$ and zero magnetic field as function
of level position $\epsilon$ at $\gamma=0.1$ (solid line) and
$\gamma=0$(dash line) }
\end{center}
\end{figure}

\begin{figure}[ht]
\begin{center}
\includegraphics [width=0.45\textwidth ]{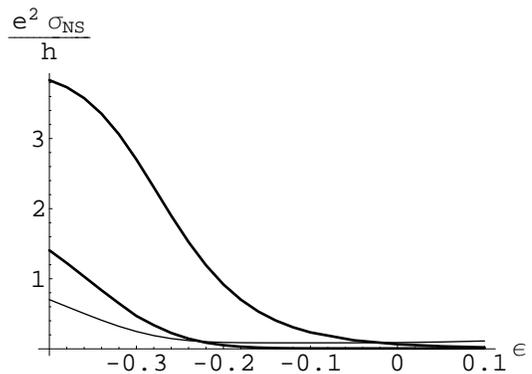}
\caption {\label{fig2} Linear conductance of an N-AB-S junction
(see Fig. \ref{fig1}) at $T=0$ and zero magnetic field as function
of level position $\epsilon$ at $\gamma=0;$ $0.1$ and
$\gamma=0.2$.}
\end{center}
\end{figure}

\begin{figure}[ht]
\begin{center}
\includegraphics[width=0.45\textwidth]{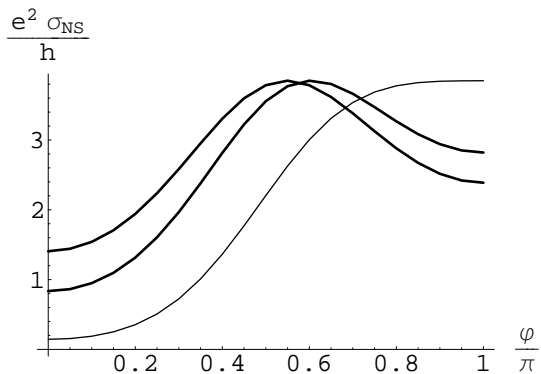} \caption
{\label{fig3} Linear conductance of an N-AB-S junction (see Fig.
\ref{fig1}) at $T=0$ versus AB phase for different level positions
($\epsilon$=-0.4 top line, -0.34 middle line, -0.24 bottom (red)
line. All curves correspond to $\gamma=0.1$. The parameters are
tuned to cover the Kondo  and crossover to Kondo regimes. }
\end{center}
\end{figure}

To obtain the zero bias differential conductance for the
noninteracting dot, expressions for the GF $ G^{R},G^{A},G^{K}$
are found for this case and are then directly substituted into
equations (\ref{cur}-\ref{G}). It is important to note that any
element in these GF which is related to the superconductor
electrode must be evaluated at energies $|\omega|<\Delta$ (this is
not the case for the self consistence equations). For a dot
modelled by a simple resonance energy level, the
 formulae for the  GF is(compare Eq.(\ref{Gm})),
 \begin{eqnarray}
G&=&\{\omega-\epsilon
\tau_{z}-\Gamma[(\tau_{z}+2\sqrt{\gamma}g_{L}F)g_
{RR}(\tau_{z}+2\sqrt{\gamma}g_{L}F^{\dagger})]\}^{-1}. \nonumber
\end{eqnarray}
Its retarded component reads then,
\begin{eqnarray}
G^{R}&=&[g+\tau_{z}g_{z}+\tau_{x}g_{x}+\tau_{y}g_{y}]^{-1},
\label{GR}
\end{eqnarray}
where $g=\omega+\Gamma\frac{i}{2}(1-P(1-\gamma))$, the
coefficients of the Pauli matrices are
$g_{z}=-\epsilon-\sqrt{\gamma}\Gamma P \cos(\varphi)$,
$g_{x}=-\frac{1}{2}\Gamma Q (1+\gamma\cos(2\varphi))$ and
$g_{y}=\frac{1}{2}\Gamma Q \gamma\sin(2\varphi)$.
 On  the Fermi surface the relations
$P=-\gamma/(1+\gamma^2)$, $Q=1/(1+\gamma^2)$ hold.
 Inserting these GF into equations ({\ref{cur}-\ref{G}}) for the current
one arrives at an expression for the zero bias conductance. After
lengthy calculations the conductance is obtained and checked to be
an even function of the flux, as is dictated by Onsager relations
(although it is not immediately apparent, as asymmetric terms
cancel).
\begin{eqnarray}
\sigma_{NS}=\frac{4 e^2 }{h}(
T_{W}+\frac{\Gamma}{2}{\cal{N}}(\varphi)\sum_{j=0}^{j=4}T_{j}\cos(j\varphi))
 \label{con}
\end{eqnarray}
Here $T_{W} =4 \gamma/(1+\gamma^2)^2$ is the background (direct)
NS transmission. Introducing the quantity
$T_{NN}=4\gamma/(1+\gamma)^2$ which is the analogous quantity
(direct transmission) for the N-AB-N junction\cite{hof} then $
T_{W} =T_{NN}^2 /(2-T_{NN})^2 .$ This is precisely the relation
between transmission coefficients for N-BB-N and N-BB-S junctions
(here BB is a ``black-box'' representing any non-interacting
scatterer) suggested in Ref.\cite{been}, using Landauer scattering
matrix approach. The coefficients $T_{W}$ and $T_{j}$ in equation
(\ref{con}) do not depend on flux. Note, however, that the
normalization factor ${\cal{N}}$ is an even function of the flux,
\begin{eqnarray}
 [{\cal{N}}(\varphi)]^{-1/2}&=&2(1+\gamma^2 )^3 [\Gamma^2
(1+\gamma)+2\epsilon^2 (1+\gamma^2) \nonumber \\
&& -4\Gamma\epsilon\gamma^{3/2}\cos(\varphi)+\Gamma^2
\gamma\cos(2\varphi)]\nonumber\\
 T_{0}&=&-8\Gamma[\Gamma^2 (-1+2\gamma^2+8\gamma^3
+5\gamma^4)+\nonumber\\
&&16\gamma\epsilon^2
(-1+\gamma-\gamma^2+\gamma^3+2\gamma^4)]\nonumber\\
 T_{1}&=&64 \sqrt{\gamma}\epsilon[\Gamma^2
(-1-2\gamma^2+\nonumber\\
&&4\gamma^3
+5\gamma^4)+4\gamma\epsilon^2 (-1+\gamma^4)]\nonumber\\
 T_{2}&=&-64\Gamma\gamma[\Gamma^2 \gamma^2(1+\gamma)+\epsilon^2
(-1+5\gamma^4)]\nonumber
\end{eqnarray}
The last two coefficients are
$T_{3}=128\sqrt{\gamma}\gamma^4\Gamma^2\epsilon$ and
$T_{4}=-16\gamma^{4}\Gamma^3$.  Here
 the result of calculations of conductance as function of
gate voltage (\ref{con}) (with $\Gamma=0.04$ and zero AB phase) is
shown on Fig.\ref{fig20}. A typical Fano asymmetry form in the
case $\gamma=0.1$ is easily detected.
As was noted above, in the MFA,
$\Gamma$ and $\epsilon$ appearing in the above equations should
respectively be replaced by $T_{K}$ and $\epsilon_{r}$, which, in
turn, are obtained through the solutions of the self-consistence
equations. The conductance of an N-AB-S junction as function of
the level position $\epsilon$ is displayed in figure \ref{fig2}
and as a function of the flux in figure \ref{fig3}.

\noindent
 {\it Discussion}: If direct tunnelling is completely
suppressed $(\gamma=0)$ the electron trajectory passes solely
through the quantum dot. This is a single channel Coulomb blockade
situation for an N-QD-S junction. For level energies which satisfy
pure Kondo limit, $ |\epsilon|>>\Gamma,$ one expects Kondo
behavior and Kondo plateau in the differential conductance at zero
bias\cite{raimondi}. We performed calculations for this range of
$\epsilon$ and found such a dependence of the linear conductance.
If the above inequality is less pronounced (as is typical for
carbon nanotubes), the Kondo effect is not completely destroyed:
The linear conductance approaches the unitary limit as
$|\epsilon|$ grows. In figure 3, depicting the conductance as
function of the level position at zero AB flux and different
background transmissions $(T_{NN}=0,0.33,0.55$, or $\gamma=0,0.1,
0.2 )$ clearly shows this behavior. In case $\gamma\neq 0$ there
is a clear suppression of the Kondo effect and the attenuation of
the conductance at its plateau step is evident although, in this
region of energies, the Kondo effect survives. This also follows
from solution of the mean field equations (for $\gamma=0.1 $)
yielding an effective Kondo temperature $T_{K}$ which is somewhat
smaller than in the case $\gamma=0$. The situation here is
reminiscent of the Fano Kondo effect in normal junctions
\cite{hof,bulka}. Such behavior of the conductance may serve as an
indication of Kondo correlations in the dot.

The crossover to the Kondo regime reveals itself more clearly in
Fig.4. It manifests a remarkable behavior of the conductance as
function of the magnetic field (the AB flux) for three different
values of the dot level position. For the two lower energies there
is a maximum very close to the Unitary limit $4e^2/h$ which is
pinned near $\varphi=\pi/2$. This property characterizes solely
the Kondo physics. It has also been proved by performing the
calculations in the strict Kondo limit $|\epsilon|>>\Gamma$. For
the higher energy in Fig.4, the QD is in the mixed valence regime
and the conductance maximum is not pinned at $\varphi=\pi/2$.
\begin{acknowledgments}
This research is partly supported by grants from the Israeli
Science Foundation (ISF) and the American Israeli Binational
Science foundation (BSF). We would like to thank Moshe Shechter
and B. Bulka for very helpful discussions.
\end{acknowledgments}

\end{document}